\documentclass[paper]{JHEP3}
\pdfoutput=1
\usepackage{amsmath,amssymb,amsthm,amscd,graphicx}
\usepackage{wrapfig}

\theoremstyle{definition}

\newcommand{\be}{\begin{equation}}
\newcommand{\ee}{\end{equation}}
\newcommand{\ba}{\begin{aligned}}
\newcommand{\ea}{\end{aligned}}
\newcommand{\bea}{\begin{eqnarray}\displaystyle}
\newcommand{\eea}{\end{eqnarray}}
\newcommand{\nn}{\nonumber}

\newdimen\tableauside\tableauside=1ex
\newdimen\tableaurule\tableaurule=0.8pt
\newdimen\tableaustep
\def\phantomhrule#1{\hbox{\vbox to0pt{\hrule height\tableaurule width#1\vss}}}
\def\phantomvrule#1{\vbox{\hbox to0pt{\vrule width\tableaurule height#1\hss}}}
\def\sqr{\vbox{%
  \phantomhrule\tableaustep
  \hbox{\phantomvrule\tableaustep\kern\tableaustep\phantomvrule\tableaustep}%
  \hbox{\vbox{\phantomhrule\tableauside}\kern-\tableaurule}}}
\def\squares#1{\hbox{\count0=#1\noindent\loop\sqr
  \advance\count0 by-1 \ifnum\count0>0\repeat}}
\def\tableau#1{\vcenter{\offinterlineskip
  \tableaustep=\tableauside\advance\tableaustep by-\tableaurule
  \kern\normallineskip\hbox
    {\kern\normallineskip\vbox
      {\gettableau#1 0 }%
     \kern\normallineskip\kern\tableaurule}%
  \kern\normallineskip\kern\tableaurule}}
\def\gettableau#1{\ifnum#1=0\let\next=\null\else
\squares{#1}\let\next=\gettableau\fi\next}

\newcommand{\figref}[1]{Fig.~\protect\ref{#1}}

\title{Refined Hopf Link Revisited}

\author{Amer Iqbal$^{\S}$, Can Koz\c{c}az${}^\dagger$
\\
$^{\S}$
 Department of Physics, LUMS School of Science \& Engineering, U-Block, D.H.A, Lahore, Pakistan.\\\\
$^{\S}$
 Department of Mathematics, LUMS School of Science \& Engineering, U-Block, D.H.A, Lahore, Pakistan.\\\\
$^\dagger$
PH-TH Division, CERN, Gen\`eve, CH-1211 Switzerland
}

\preprint{CERN-2011/267}
\abstract{We establish a relation between the refined Hopf link invariant and the S-matrix of the refined Chern-Simons theory. We show that the refined open string partition function corresponding to the Hopf link, calculated using the refined topological vertex, when expressed in the basis of Macdonald polynomials gives the S-matrix of the refined Chern-Simons theory.}

\begin{document}

\section{Introduction}
The connection between the topological classification of knots and links with the topological field theories goes back to the work by Witten realizing the knot polynomials as expectation values of Wilson loops in Chern-Simons theory \cite{Witten:1988hf}. Although this connection defined a three dimensional definition of the Jones polynomials it did not answer questions about the integrality of the coefficients appearing in the polynomials. The interpretations for these integers has been offered by both the mathematicians and physicists. On the mathematics side, a doubly graded homology is associated to each knot \cite{Khovanov}. The Euler characteristic of the homology complex was shown to give the knot polynomials. The coefficients appear as the dimension of the vector spaces in the complex, hence, the positivity and integrality become manifest. The physics answer was given in the terms of counting BPS states. In a type IIA compactification on Calabi-Yau threefold, the relevant BPS states come from D2-branes wrapping two cycles with boundaries on Lagrangian cycles around which some D4-branes are wrapped and extend in the non-compact transverse space \cite{Ooguri:1999bv}.

The Poincar\'{e} polynomial of the doubly graded complex includes more information about the knot and depends on one more parameter than the corresponding Euler characteristic. It did not take a long time to realize that these more detailed polynomials can be achieved in terms of a more refined counting of the BPS states in M-theory compactifications \cite{Gukov:2004hz}. The refinement was motivated by the equivariant computation, with respect to the $\mathbb{C}^{\times}\times \mathbb{C}^{\times}$ action on $\mathbb{C}^2$ given by $(z_{1},z_{2})\mapsto (e^{\epsilon_{1}}z_{1},e^{\epsilon_2}z_{2})$, of instantons in ${\cal N}=2$ supersymmetric gauge theories \cite{Nekrasov:2002qd} and was related to BPS invariants in \cite{Hollowood:2003cv} for the case $\epsilon_1+\epsilon_2\neq 0$. The formalism to compute these refined counting of BPS states within the topological string theory on toric Calabi-Yau threefolds was developed in \cite{Iqbal:2007ii}. Later it was shown that the refined knot polynomials can be calculated using the refined topological vertex in \cite{Gukov:2007tf}. These results hinted at the existence of some refinement of the Chern-Simons theory incorporating the extra deformation $\epsilon_{1}+\epsilon_{2}$. Such a refinement of the Chern-Simons theory for 3-manifolds with circle action (Seifert spaces) was recently constructed in \cite{Aganagic:2011sg}.

In this short note we revisit the refined Hopf link studied in \cite{Gukov:2007tf}. In an attempt to understand the relation between the refined Hopf link and the refined Chern-Simons theory we find that the open string partition function calculated from the refined topological vertex seems to be written in a mixed basis of Schur and Macdonald polynomials with Macdonald polynomials associated with the preferred direction. The fact that the preferred direction of the refined topological vertex seems to have the holonomy which gives Macdonald polynomials rather than Schur polynomials was also indicated in \cite{Kozcaz:2010af, Dimofte:2010tz}. We argue that this is indeed the case by looking at the structure of the free energy for the case of the unknot. With a choice of basis of symmetric functions in mind we calculate the Hopf link partition function and show that when written in the basis of Macdonald polynomials the coefficients are the elements of the S-matrix of the refined Chern-Simons theory. Thus in this case the refined topological vertex and the refined Chern-Simons theory calculation give the same open string partition function. We also show that, up to framing factors which need to be studied properly, the two different choices for the preferred direction give the same open string partition function. This seems to indicate that open string partition functions may also be independent of the preferred direction, as it was argued for closed amplitudes in \cite{Iqbal:2007ii} and was used to develop new identities in symmetric functions in \cite{Iqbal:2008ra}.

The paper is organized as follows. In section 2 we recall some results of Etingof and Kirillov which are related to the refined Chern-Simons theory and use the idea of modular tensor categories to establish an important identity which is used is section 3. In section 3 we argue the preferred direction of the refined topological vertex should be associated with Macdonald polynomial rather than Schur polynomial. In section 4 we calculate the open string partition function for the Hopf link and show when expressed in terms of Macdonald polynomials it gives the S-matrix of the refined Chern-Simons theory. The calculation of the refined Hopf link in this section relies heavily on the results of \cite{Awata:2009sz}. We also show that for the brane configuration corresponding to the Hopf link the open string partition function is also invariant under a change in the choice of the preferred direction.

\section{Modular Tensor Categories and Refined Chern-Simons}

In \cite{Aganagic:2011sg} a refinement of the Chern-Simons theory based on a supersymmetric index was described. The theory was engineered using M5-branes wrapped on a 3-manifold, inside the cotangent space of the 3-manifold, with M2-branes ending on Lagrangian submanifold providing the knots and links in the 3-manifold \cite{Ooguri:1999bv}. Apart from the usual parameter $q=\mbox{exp}\Big(\frac{2\pi i}{k+N}\Big)$ which is present in the $SU(N)$ Chern-Simons theory with coupling constant $k$, the refinement introduces a new parameter $t$ which appears in the supersymmetric index. The refinement requires an extra $U(1)$ symmetry for the 3-manifold and therefore the refined Chern-Simons theory only exists for Seifert manifolds.

It was argued in \cite{Aganagic:2011sg} that the quantization of this theory on $M$, a solid torus with boundary $T^{2}$, gives a a Hilbert space which is again labelled by integrable highest weight representation of affine $SU(N)$ at level $k$ and the wavefunction on a solid torus with Wilson line in representation $\lambda$ is given by
\bea
Z_{\lambda}(M)=\Delta_{q,t}\,P_{\lambda}({\bf x};q,t),
\eea
where ${\bf x}$ are the eigenvalues of the holonomy of the flat connection on the a-cycle of the boundary $T^{2}$, $P_{\lambda}({\bf x};q,t)$ is the Macdonald polynomial \cite{macdonald} and
\bea
\Delta_{q,t}=\prod_{m=0}^{\beta-1}\prod_{1\leq i<j\leq N}\Big(q^{\frac{m}{2}}\,x_{i}-q^{-\frac{m}{2}}\,x_{j}\Big)\,,\,\,\,\,\,\,\,\,t=q^{\beta}\,,\,\,\beta\in \mathbb{Z}_{\geq 0}\,.
\eea
It was argued that the fusion coefficients $\widehat{N}^{\nu}_{\lambda\,\mu}$ are not integers any more but are rational functions of $q$ and $t$ given by
\bea
P_{\lambda}({\bf x};q,t)\,P_{\mu}({\bf x};q,t)=\sum_{\nu}\widehat{N}^{\nu}_{\lambda\,\mu}P_{\nu}({\bf x};q,t)\,.
\eea
The S-matrix of the theory is a deformation of the usual S-matrix of the G/G WZW theory ($G=SU(N)$) and is given by
\bea
S_{\lambda\,\mu}=P_{\lambda}(t^{\frac{1}{2}},t^{\frac{3}{2}},\mathellipsis,t^{N-\frac{1}{2}};q,t)\,P_{\mu}(t^{\frac{1}{2}}q^{-\lambda_{1}},t^{\frac{3}{2}}q^{-\lambda_{2}},\mathellipsis,t^{N-\frac{1}{2}}q^{-\lambda_{N}};q,t).
\eea
It is interesting to note that the above S-matrix was studied by Etingof and Kirillov \cite{Kirillov} in the context of a modular tensor category based on the quantum group $U_{q}(sl_{N})$ with $q$ being a root of unity.

The idea of modular tensor categories appeared in the study of rational conformal field theories and it is the structure underlying a topological quantum field theory in 3-dimensions. They first appeared in the work of Moore and Seiberg \cite{Moore:1988uz,Moore:1988qv}.  The essential idea is that there are a finite set of objects associated with a two dimensional surface. The topological nature of the association implies that the mapping class group of the surface acts on these objects. Perhaps the most well known example is the one coming from the $G/G$ WZW theory on $T^{2}$ for $G=SU(N)$. In this case the objects are the conformal blocks of the theory which are the characters of $\widehat{SU(N)}_{k}$, the affine Lie algebra of $SU(N)$ at level $k$.
The action of the modular group on the characters of $\widehat{SU(N)}_{k}$ is well known and is given by:
\bea
S_{\lambda\,\mu}&=&\sum_{w\in W}(-1)^{|w|}\,\mbox{exp}\Big(\frac{i\pi }{k+N}(\lambda+\rho,w(\mu+\rho))\Big)\\\nn\label{xx}
&=&s_{\lambda}(q^{\frac{1}{2}},q^{\frac{3}{2}},\mathellipsis,q^{N-\frac{1}{2}})\,s_{\mu}(q^{\frac{1}{2}-\lambda_{1}},q^{\frac{3}{2}-\lambda_{2}},\mathellipsis,q^{N-\frac{1}{2}-\lambda_{N}}).
\eea
Since the WZW theory on $T^{2}$ is related to the canonical quantization of the Chern-Simons theory on $T^{2}\times \mathbb{R}$ the space of conformal blocks of the WZW theory on $T^2$ is also the Hilbert space of the Chern-Simons theory and these characters are the wavefunction of the Chen-Simons theory on solid torus with a Wilson line \cite{Elitzur:1989nr}.  In the Chern-Simons theory the normalized S-matrix $S_{\lambda\,\mu}/S_{\emptyset\,\emptyset}$ is the Hopf link invariant.

A more general construction on surfaces with marked points can also be realized. Recall that the wavefunctions of the Chern-Simons theory (conformal blocks of the WZW theory) can be obtained by path integral on the solid torus with boundary $T^{2}$ \cite{Elitzur:1989nr}. These can be realized as trace over the integrable highest weight representation. Let $V_{\lambda+\rho}$ be the finite dimensional irreducible representation of $G=SU(N)$ with highest weight $\lambda$, then
\bea
\psi({\bf x})=\mbox{Tr}_{V_{\rho}}(e^{\vec{\bf x}\cdot \vec{h}})=\prod_{1\leq i<j\leq N}(e^{x_{i}}-e^{x_{j}}),
\eea
where $\vec{h}$ are the Cartan generators. This is the partition function of the Chern-Simons theory on the solid torus with $e^{x_{i}}$ being the eigenvalues of the holonomy of the flat connection around the $a$ cycle of the boundary \cite{Elitzur:1989nr}. If we insert some Wilson line on the boundary $T^{2}$ in representation with highest weight vector $\lambda$ then
\bea
\psi_{\lambda}({\bf x})=\mbox{Tr}_{V_{\lambda+\rho}}(e^{\vec{\bf x}\cdot \vec{h}})=\Big(\prod_{1\leq i<j\leq N}(e^{x_{i}}-e^{x_{j}})\Big)\,s_{\lambda}(e^{x_{1}},e^{x_{2}},\mathellipsis,e^{x_{N}}),
\eea
where $s_{\lambda}$ is the Schur polynomial \cite{macdonald}.

In \cite{Kirillov, Kirillov2} the above construction was generalized to include a marked point, with holonomy in the representation $U$, on the torus.  Let $U = S^{(\beta-1)\,N}\mathbb{C}^{N}$, $(\beta-1)N$-th symmetric representation of $SU(N)$. Let $\Phi$ be a non-zero intertwining operator $V \mapsto V \otimes U$ which exists if $\lambda\geq
(\beta-1)\,\rho$. Let $\lambda = \mu + (\beta-1)\,\rho$, where $\mu$ is any dominant integral weight then \cite{Kirillov}
\bea
\psi^{\beta}_{\lambda}=\mbox{Tr}|_{V_{\lambda+\rho}} (\Phi\, e^{\vec{\bf x}\cdot \vec{h}})=\Big(\prod_{1\leq i<j\leq N}(e^{x_{i}}-e^{x_{j}})\Big)^{\beta}\,J_{\lambda}^{(\beta)}(e^{x_{1}},e^{x_{2}},\mathellipsis,e^{x_{N}}),
\eea where $J_{\lambda}^{(\alpha)}$ are Jack symmetric functions \cite{macdonald}. This corresponds to the beta-deformation of the Vandermonde measure. The generalization to Macdonald polynomials requires working with quantum group modules \cite{Kirillov, Kirillov2}.  Let $V_{\lambda+\rho}$ and $S^{(\beta-1)\,N}\mathbb{C}^{N}$ be the $q$-deformation of the $SU(N)$-modules so that these are now $U_{q}(sl_N)$ modules. Then \cite{Kirillov2}
\bea
\psi^{\beta}_{\lambda}=Tr|_{V_{\lambda}}(\Phi\,e^{\vec{x}\cdot h})=\Delta_{q,t}(x)\,P_{\lambda}(x;q,q^{\beta}).
\eea
In this case the states admit an $PSL(2,\mathbb{Z})$ action for which the S-matrix is given by
\bea\label{s2}
S_{\lambda\,\mu}=P_{\lambda}(t^{\frac{1}{2}},t^{\frac{3}{2}},\mathellipsis,t^{N-\frac{1}{2}};q,t)\,P_{\mu}(t^{\frac{1}{2}}q^{-\lambda_{1}},t^{\frac{3}{2}}q^{-\lambda_{2}},\mathellipsis,t^{N-\frac{1}{2}}q^{-\lambda_{N}};q,t).
\eea
This is precisely the S-matrix of the refined Chern-Simons theory \cite{Aganagic:2011sg}.

An alternative expression for the S-matrix follows from the axioms of the modular tensor category \cite{Kirillov3,Rowell} and  is given by the fusion coefficients $N^{\nu}_{\lambda\,\mu}$ defined using the simple objects in the category
\bea
V_{\lambda}\otimes V_{\mu}=\oplus_{\nu}\,N^{\nu}_{\lambda\,\mu}\,V_{\nu}\,,
\eea
twists $\theta_{\lambda}:V_{\lambda}\mapsto V_{\lambda}$ satisfying certain conditions \cite{Kirillov3} and the dimension of the simple object in the category $\mbox{dim}\,V_{\lambda}$,
\bea\label{s}
S_{\lambda\,\mu}=\theta_{\lambda}^{-1}\theta_{\mu}^{-1}\sum_{\nu}N^{\nu}_{\lambda\,\mu}\,\theta_{\nu}\,\mbox{dim}V_{\nu}.
\eea
In the case of the category based on the integrable representations of $\widehat{SU(N)}_{k}$ the fusion coefficients are the Littlewood-Richardson coefficients, the twists $\theta_{\lambda}=q^{\langle\lambda,\lambda+\rho\rangle}$ are the framing factors and the dimension of the objects in the category are the quantum dimension of the corresponding integrable representation,
\bea
S_{\lambda\,\mu}&=&q^{\kappa(\lambda)+\kappa(\mu)}\,\sum_{\nu}\,N^{\nu}_{\lambda\,\mu}\,q^{-\kappa(\nu)}\,\mbox{dim}_{q}V_{\nu}\\\nn
&=&q^{\kappa(\lambda)+\kappa(\mu)}\,\sum_{\nu}\,N^{\nu}_{\lambda\,\mu}\,q^{-\kappa(\nu)}\,s_{\nu}(q^{\frac{1}{2}},q^{\frac{3}{2}},\mathellipsis q^{N-\frac{1}{2}}).\\\nn
\eea
The fact that the above expression for $S_{\lambda\,\mu}$ is equal to the one given in Eq. (\ref{xx}) follows from a non-trivial identity which we will discuss later.

In the case of the modular tensor category based on $U_{q}(sl_N)$ the objects in the category are the irreducible highest weight modules of $U_{q}(sl_{N})$, the fusion coefficient are the fusion coefficients of the Macdonald polynomials $\widehat{N}_{\lambda\,\mu}^{\nu}$ and the twists are given by $\theta_{\lambda}$ are given by the framing factors $f_{\lambda}=t^{\Arrowvert\eta^t\Arrowvert^2/2}\,q^{-\Arrowvert\eta\Arrowvert^2/2}$ and the dimensions of the highest weight modules are give by principal specialization of Macdonald polynomials $P_{\lambda}(t^{\frac{1}{2}},\mathellipsis,t^{N-\frac{1}{2}};q,t)$. In this case the Eq. (\ref{s}) gives an alternative representation of the S-matrix given by
\bea\label{s3}
S_{\lambda\,\mu}=\frac{1}{f_{\lambda}\,f_{\mu}}\,\sum_{\nu}\widehat{N}^{\nu}_{\lambda\,\mu}\,f_{\nu}\,P_{\nu}(t^{\frac{1}{2}},\mathellipsis,t^{N-\frac{1}{2}};q,t)\,.
\eea
The fact that Eq. (\ref{s2}) and Eq. (\ref{s3}) are equal is a non trivial identity
\bea\label{identityx}
\hskip -1.0cm P_{\lambda}(t^{\frac{1}{2}},\cdot\cdot,t^{N-\frac{1}{2}};q,t)\,P_{\mu}(t^{\frac{1}{2}}q^{-\lambda_{1}},\cdot\cdot,t^{N-\frac{1}{2}}q^{-\lambda_{N}};q,t)=\frac{1}{f_{\lambda}\,f_{\mu}}\,\sum_{\nu}\widehat{N}^{\nu}_{\lambda\,\mu}\,f_{\nu}\,&& P_{\nu}(t^{\frac{1}{2}},\cdot\cdot,t^{N-\frac{1}{2}};q,t)
\eea
which is proved in appendix B.  We will encounter Eq. (\ref{s3}) again while studying the open string partition function corresponding to the Hopf link invariant. The left hand side of the above identity is what appears naturally from the open string partition function, corresponding to the Hopf link calculated using the refined topological vertex, and the right hand side is what appears in the refined Chern-Simons theory.

\section{Open String Amplitudes and Basis  of Symmetric Functions}
In this section we will discuss the open string partition function for certain configuration of branes on Lagrangian cycles in $\mathbb{C}^3$. We will use the refined topological vertex formalism to calculate the open string partition functions. We will argue that for branes ending on the preferred direction of the refined topological vertex the open string partition function should be expanded in terms of Macdonald polynomials rather then Schur polynomials.

\begin{figure}[h]
\begin{center}
\includegraphics[scale=0.7]{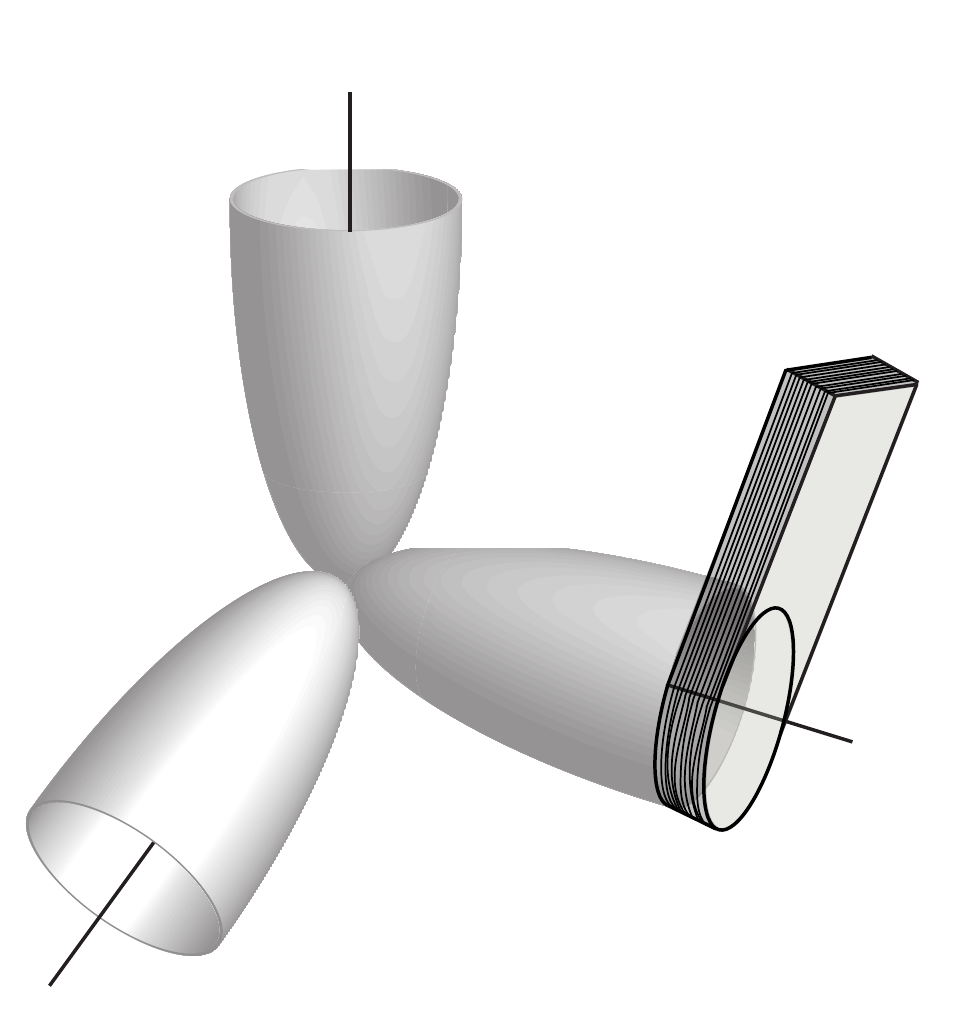}
\end{center}
\caption{$\mathbb{C}^{3}$ with a stack of D-branes on one of the non-compact directions}\label{fig1}
\end{figure}

The open topological string partition function captures the invariants of holomorphic curves with boundaries on the Lagrangian cycles in a Calabi-Yau threefold. Physically these invariants are the BPS degeneracies of the D2-branes wrapping holomorphic curves and ending on the D4-branes which are wrapping the Lagrangian cycles in the internal space and filling $\mathbb{R}^{2}\subset\mathbb{R}^{4}$ in the transverse space. If we consider a single Lagrangian cycle, this partition function is given by \cite{Ooguri:1999bv}
\bea
Z_{open}&=&\mbox{exp}(F_{open})\\\nn
F_{open}&=&\sum_{n=1}^{\infty}\sum_{Q,R,s}\,N_{R,Q,s}q^{\,n\,s}\,(q^{\frac{n}{2}}-q^{-\frac{n}{2}})^{-1}\lambda^{n\,Q}\,\mbox{Tr}_{R}\,U^{n},
\eea
where $U$ is the holonomy of the gauge field on the D4-brane over the boundary of the D2-brane and $Q$ is the (relative) homology of the class of the curve on which the D2-brane wraps. $N_{R,Q,s}$ are integers giving the BPS degeneracies in the transverse two dimensional theory on the D4-brane. The BPS particles are created by the fields which carry two dimensional spin $s$.  The degeneracies depend on the representation $R$ encoding different ways D2-branes can end on D4-branes. This description can be lifted to M-theory. The M5-branes wrap the Lagrangian submanifold and $\mathbb{R}^{3}$. The M2-branes can end on M5-branes and give rise to BPS particles in the non-compact three dimensional space. The little group of massive particles in three dimensions is $SO(2)$ and these particles have spin $s$ under this group. The previous construction is just the reduction on the M-theory circle. For the case of a stack of branes on $\mathbb{C}^3$, as shown in \figref{fig1}, the partition function is given by
\bea
Z_{open}&=&\mbox{exp}\Big(\sum_{n=1}^{\infty}\frac{1}{q^{\frac{n}{2}}-q^{-\frac{n}{2}}}\,\mbox{Tr}\,U^{n}\Big)\\\nn
&=&\prod_{i,j}\Big(1-x_{i}\,q^{j-\frac{1}{2}}\Big),
\eea
where $\{x_{1},x_{2},x_{3},\mathellipsis\}$ are the eigenvalues of the holonomy $U$.

A refinement of the open string partition function was given in \cite{Gukov:2004hz} and was conjectured to be related to the Khovanov homology of knots and links. The refined open partition function has one extra parameter $t$ in it (for $t=q$ it reduces to the usual open string partition function) and, for a single Lagrangian cycle, is given by
\bea
Z'_{open}=\mbox{exp}\Big( \sum_{n=1}^{\infty}\sum_{Q,R,s_{1},s_{2}}\,N_{R,Q,s_{1},s_{2}}q^{\,n\,s_{1}}\,t^{-n\,s_{2}}\,(q^{\frac{n}{2}}-q^{-\frac{n}{2}})^{-1}\lambda^{n\,Q}\,\mbox{Tr}_{R}\,U^{n}\Big).
\eea

The refinement includes a new parameter which captures an additional charge of the embedding of the D4-branes. The refinement, like in the case of closed strings, can be best understood within the M-theory. The BPS particles are charged under the little group of massive particles $SO(4)\simeq SU(2)_{L}\times SU(2)_{R}$. The spin of the particles which coupled to the parameter $q$ corresponds to $U(1)_{L}\in SU(2)_{L}$. With the new parameter we measure the $U(1)_{R}\in SU(2)_{R}$ charge.

Consider the configuration of branes shown in the \figref{fig1}. The branes are placed on the $U(1)$ invariant leg of $\mathbb{C}^3$ corresponding to the preferred direction of the refined topological vertex. The open string partition, from the refined vertex, is given by
\bea
Z_{\nu}=(-1)^{\nu}q^{\frac{\Arrowvert\nu\Arrowvert^{2}}{2}}t^{-\frac{\Arrowvert\nu^{t}\Arrowvert^{2}}{2}}C_{\emptyset\,\emptyset\,\nu^{t}}(q,t)=(-1)^{|\nu|}P_{\nu^t}(q^{-\rho};t,q)\,.
\eea
If we assume the trace of the holonomy to be given by the Schur polynomial then
\bea
Z(U)&=&\sum_{\nu}\,Z_{\nu}\,\mbox{Tr}_{\nu}U\\\nn
&=&\sum_{\nu}(-1)^{|\nu|}\,P_{\nu^t}(q^{-\rho};t,q)\,s_{\nu}({\bf x}).
\eea
The free energy is given by
\bea
F_{open}&=&\mbox{log}\,Z(U)\\\nn
&=&\frac{1}{q^{\frac{1}{2}}-q^{-\frac{1}{2}}}\,\mbox{Tr}_{\tableau{1}}\,U+\frac{1}{2(q-q^{-1})}\mbox{Tr}_{\tableau{1}}U^2+\frac{q(t-q)}{2(1-q)(q-q^{-1})(1-q\,t)}\,Tr_{\tableau{2}}\,U+\mathellipsis.
\eea
It is clear that the above expansion does not match the general form of the open string amplitudes for $t\neq q$. The partition function obtained by putting a brane on the preferred direction should not be different from the one obtained by putting the brane on an un-preferred leg since both get contribution from the single holomorphic disk. Consider the following partition function
\bea
Z'(U)&=&\sum_{\nu}(-1)^{|\nu|}\,P_{\nu^t}(q^{-\rho};t,q)\,P_{\nu}({\bf x};q,t)\\\nn
&=&\prod_{i,j}\Big(1-x_{i}\,q^{i-\frac{1}{2}}\Big)=\mbox{exp}\Big(\sum_{n=1}^{\infty}\frac{1}{q^{\frac{n}{2}}-q^{-\frac{n}{2}}}\mbox{Tr}\,U^n\Big).
\eea
This is exactly the open string partition function with branes on an un-preferred direction (and the same as the one calculated using the topological vertex). This suggests that the refined topological vertex, which was defined combinatorially as counting anisotropic plane partitions, is in written in a basis in which the preferred direction is associated with Macdonald polynomial. The partition function of $\mathbb{C}^{3}$ is thus given by
\bea
Z(U_{1},U_{2},U_{3})=\sum_{\lambda\,\mu\,\nu}\,C_{\lambda\,\mu\,\nu}(t,q)\,s_{\lambda}(U_{1})\,s_{\mu}(U_{2})\,P_{\nu}(U_{3};t,q).
\eea
The above partition function is a symmetric function of the eigenvalues of $U_{1}$, $U_{2}$ and $U_{3}$ and therefore can be expressed in any basis of symmetric functions. We will show in the next section, using two examples, that open topological string partition functions calculated using the refined vertex when expressed in the appropriate basis are independent of the choice of preferred direction and that the S-matrix of the refined Chern-Simons theory is equal to the refined partition function of the Hopf link \cite{Gukov:2007tf} when expressed in the basis of Macdonald polynomials.

\section{S-matrix of the Refined Topological Vertex}
In this section we will calculate the S-matrix of the refined topological vertex (refined Hopf link) and show that it gives the S-matrix of the refined Chern-Simons theory.

\par{In \cite{Gukov:2007tf}, the connection between the homological knot invariants of the colored Hopf link and the
refined topological vertex was established. In this section we want to review this computation before exploring other
possible brane configurations.}
\par{We compute the open topological string amplitudes on the resolved conifold with two branes on the external legs as depicted in \figref{branes}.
\begin{figure}[h]
\begin{center}
\includegraphics[scale=0.5]{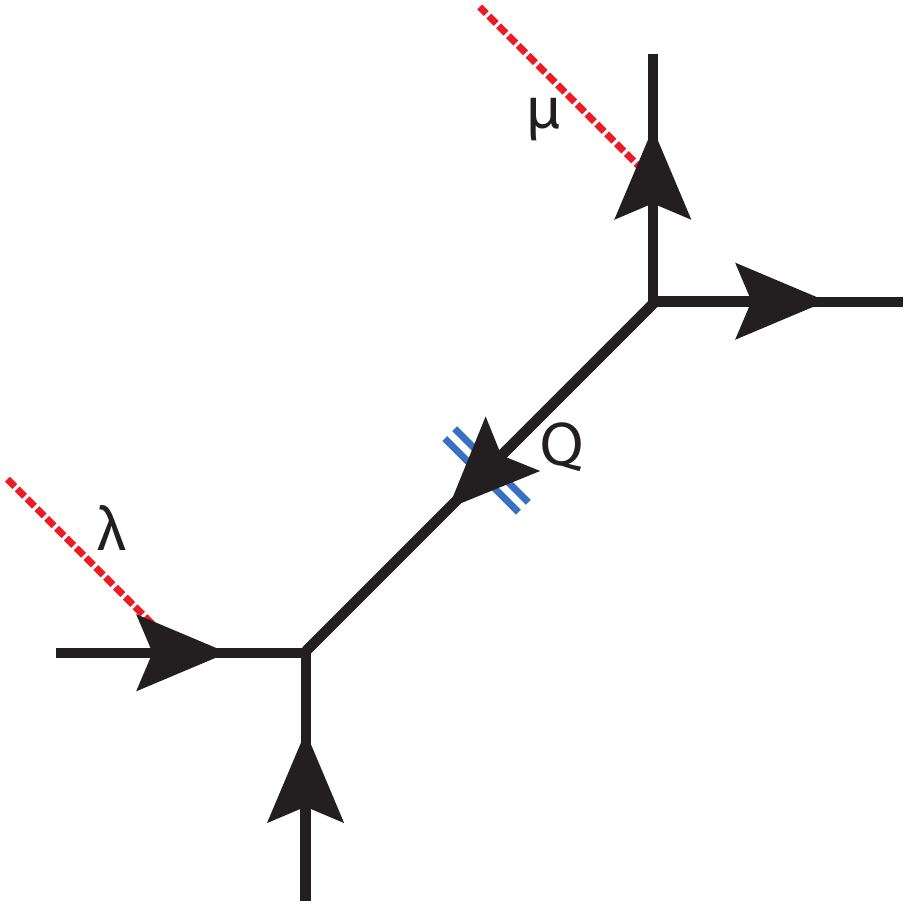}
\end{center}
\caption{The brane configuration giving rise to Hopf link}\label{branes}
\end{figure}
The preferred direction chosen to be along the internal leg. According to the rules of the refined topological vertex the open amplitude is given by}
\begin{align}\nn\label{pf}
Z_{\lambda\mu}&=\sum_{\nu}(-Q)^{|\nu|}\, C_{\emptyset\lambda^{t}\nu^{t}}(q,t) C_{\mu\emptyset\nu}(t,q)\\
&=\left(\frac{q}{t}\right)^{\frac{|\mu|}{2}}t^{n(\lambda^{t})}q^{-n(\lambda)}\sum_{\nu}(-Q)^{|\nu|}P_{\nu}(t^{-\rho};q,t)P_{\nu^{t}}(q^{-\rho};t,q)s_{\lambda^t}(q^{-\nu}t^{-\rho})s_{\mu^t}(q^{-\nu}t^{-\rho}).
\end{align}
If we introduce one unit of framing for the brane on the horizontal leg (with representation $\lambda$) then we get
\bea
Z^{(1,0)}_{\lambda\,\mu}=\left(\frac{q}{t}\right)^{\frac{|\lambda|+|\mu|}{2}}\sum_{\nu}(-Q)^{|\nu|}P_{\nu}(t^{-\rho};q,t)P_{\nu^{t}}(q^{-\rho};t,q)s_{\lambda^t}(q^{-\nu}t^{-\rho})s_{\mu^t}(q^{-\nu}t^{-\rho}).
\eea
The superscript $(1,0)$ denotes the framing for the two branes. Let $U$ and $V$ be the holonomies for the two branes, the open string partition function is then given by
\bea\nn
Z(U,V)&=&\sum_{\lambda\,\mu}Z^{(1,0)}_{\lambda\,\mu}\,\mbox{Tr}_{\lambda}U\,\mbox{Tr}_{\nu}V\\\nn
&=&\sum_{\lambda\,\mu}\left(\frac{q}{t}\right)^{\frac{|\lambda|+|\mu|}{2}}\sum_{\nu}(-Q)^{|\nu|}P_{\nu}(t^{-\rho};q,t)P_{\nu^{t}}(q^{-\rho};t,q)s_{\lambda^t}(q^{-\nu}t^{-\rho})s_{\mu^t}(q^{-\nu}t^{-\rho})\,s_{\lambda}({\bf x})\,s_{\mu}({\bf y})\\\nn
&=&\sum_{\nu}(-Q)^{|\nu|}P_{\nu}(t^{-\rho};q,t)P_{\nu^{t}}(q^{-\rho};t,q)\prod_{i,j}(1+x_{i}\,\sqrt{\frac{q}{t}}\,t^{-\rho_{i}}q^{-\nu_{i}})\,\prod_{i,j}(1+y_{i}\,\sqrt{\frac{q}{t}}\,t^{-\rho_{i}}q^{-\nu_{i}})\,,
\eea
where $\{x_{1},x_{2},\mathellipsis\}$ and $\{y_{1},y_{2},\mathellipsis\}$ are eigenvalues of $U$ and $V$ respectively. $Z(U,V)$ is a symmetric function of $\{x_{1},x_{2},\mathellipsis\}$ and $\{y_{1},y_{2},\mathellipsis\}$ and if we express it in terms of Macdonald polynomials rather than Schur polynomials we get
\bea\nn
\hskip -0.9cmZ(U,V)&=&\sum_{\lambda\,\mu}\widetilde{Z}^{(1,0)}_{\lambda\,\mu}\,P_{\lambda}({\bf x};t,q)\,P_{\mu}({\bf y};t,q)\\\label{xcv}
\hskip -0.9cm\widetilde{Z}^{(1,0)}_{\lambda\,\mu}&=&\left(\frac{q}{t}\right)^{\frac{|\lambda|+|\mu|}{2}}\sum_{\nu}(-Q)^{|\nu|}P_{\nu}(t^{-\rho};q,t)P_{\nu^{t}}(q^{-\rho};t,q)P_{\lambda^t}(q^{-\nu}t^{-\rho};q,t)P_{\mu^t}(q^{-\nu}t^{-\rho};q,t)\,.
\eea

\par{Although the open topological string amplitude is an infinite expansion in $Q$ when we normalize it by the closed string partition function it becomes a polynomial of degree $|\lambda|+|\mu|$ \cite{Gukov:2007tf}. This was proved in \cite{Awata:2009sz} and we will closely follow their computation with slight change in appendix C. The normalized open amplitude takes the following form}
\begin{align}\label{norm1}\nonumber
Z_{\lambda\,\mu}^{\mbox{\tiny{norm}}}&\equiv \frac{\widetilde{Z}^{(1,0)}_{\lambda\,\mu}}{Z_{\emptyset\,\emptyset}}=\left(\frac{q}{t}\right)^{\frac{|\lambda|+|\mu|}{2}}\sum_{\sigma}\widehat{N}^{\sigma}_{\lambda^t\,\mu^t}\,P_{\sigma}(t^{-\rho};q,t)\,\prod_{(i,j)\in \sigma}(1-Q\,t^{i-\frac{1}{2}}\,q^{-j+\frac{1}{2}})\\
&=\left(-Q\sqrt{\frac{q}{t}}\right)^{|\lambda|+|\mu|}\sum_{\sigma}\widehat{N}^{\sigma}_{\lambda^t\,\mu^t}\, t^{\frac{\Arrowvert\sigma^{t}\Arrowvert^{2}}{2}}q^{-\frac{\Arrowvert\sigma\Arrowvert^{2}}{2}}P_{\sigma}\Big(t^{-\rho},Q^{-1}\sqrt{\frac{q}{t}}\,t^{\rho};q,t\Big),
\end{align}
where $\widehat{N}^{\eta}_{\lambda\,\mu}$ are the analog of Littlewood-Richardson coefficients for the Macdonald polynomials,
\bea
P_{\lambda}({\bf x};q,t)\,P_{\mu}({\bf y};q,t)=\sum_{\sigma}\,\widehat{N}^{\sigma}_{\lambda\,\mu}\,P_{\sigma}({\bf x};q,t)\,.
\eea
$\widehat{N}^{\sigma}_{\lambda\,\mu}$ are rational functions of $t$ and $q$ with the property that $\widehat{N}^{\sigma}_{\lambda\,\mu}=0$ if $|\sigma|\neq |\lambda|+|\mu|$ and $\widehat{N}^{\sigma}_{\lambda\,\emptyset}=\delta_{\sigma\,\lambda}$:
\bea
\widehat{N}^{\tableau{2}}_{\tableau{1}\,\tableau{1}}&=&1\,,\,\,\,\,\widehat{N}^{\tableau{1 1}}_{\tableau{1}\,\tableau{1}}=\frac{(1-t)(1+q)}{1-t\,q}\\\nn
\widehat{N}^{\tableau{3}}_{\tableau{2}\,\tableau{1}}&=&1\,,\,\,\,\,\widehat{N}^{\tableau{2 1}}_{\tableau{2}\,\tableau{1}}=\frac{(1-q\,t^2)(1-q^2)}{(1-t\,q)(1-t\,q^2)}\,,\,\,\,\,\,\widehat{N}^{\tableau{1 1 1}}_{\tableau{2}\,\tableau{1}}=0\,.
\eea

\par{The specialization of the Macdonald polynomial given in Eq. (\ref{norm1}) has a very important property. To see this note that the Macdonald polynomials are symmetric functions and, hence, can be expressed in terms of another basis of the space of symmetric functions: the power sum symmetric functions $p_{k}({\bf x})=\sum_{i}x_{i}^{k}$. Consider the power sum symmetric functions for ${\bf x}=\{Q^{-1}\sqrt{\frac{q}{t}}\,t^{\rho},\,t^{-\rho}\,q^{-\sigma}\}$ and lets take $Q^{-1}\sqrt{\frac{q}{t}}=t^{N}$ ($N\geq\ell(\sigma$)),
\bea
p_{k}\Big(Q^{-1}\sqrt{\frac{q}{t}}\,t^{\rho},\,t^{-\rho}\,q^{-\sigma}\Big)&=&-t^{Nk}\,\frac{t^\frac{k}{2}}{1-t^{k}}+\sum_{i=1}^{N}\,t^{k(i-\frac{1}{2})}\,q^{-k\sigma_{i}}\,+\frac{t^{k(N+\frac{1}{2})}}{1-t^k}\\\nn
&=&\sum_{i=1}^{N}\,t^{k(i-\frac{1}{2})}\,q^{-k\sigma_{i}}\,.
\eea
where we used analytic continuation for the geometric sums in $t$. Thus the Macdonald polynomial for these variables is actually a polynomial in $t$ and $q$. Thus for $Q^{-1}\sqrt{\frac{q}{t}}=t^{N}$ the normalized partition functions is given by
\bea
Z_{\lambda\,\mu}^{\mbox{\tiny{norm}}}=\left(-\frac{q}{t^{N+1}}\right)^{|\lambda|+|\mu|}\sum_{\sigma}\widehat{N}^{\sigma}_{\lambda^t\,\mu^t}\, t^{\frac{\Arrowvert\sigma^{t}\Arrowvert^{2}}{2}}q^{-\frac{\Arrowvert\sigma\Arrowvert^{2}}{2}}
P_{\sigma}\Big(t^{\frac{1}{2}},t^{\frac{3}{2}},\mathellipsis t^{N-\frac{1}{2}};q,t\Big)\,.
\eea
Using identity Eq. (\ref{identityx}) (see appendix B for proof)
\bea
\frac{1}{f_{\lambda}\,f_{\mu}}\,\sum_{\sigma}\widehat{N}^{\sigma}_{\lambda\,\mu}\, f_{\sigma}\,P_{\sigma}(t^{-\rho},z\,t^{\rho};q,t)=P_{\lambda}(t^{-\rho},z\,t^{\rho};q,t)\,P_{\mu}(t^{-\rho}q^{-\lambda},z\,t^{\rho};q,t)\,,
\eea
where $f_{\sigma}= t^{\frac{\Arrowvert\sigma^{t}\Arrowvert^{2}}{2}}q^{-\frac{\Arrowvert\sigma\Arrowvert^{2}}{2}}$ we get ($N\geq \ell(\lambda^t),\ell(\mu^t)$)
\bea\label{internal}
\hskip -0.8cm Z_{\lambda\,\mu}^{\mbox{\tiny{norm}}}=\left(-\frac{q}{t^{N+1}}\right)^{|\lambda|+|\mu|}\,f_{\lambda^t}\,f_{\mu^t}\,P_{\lambda^t}(t^{\frac{1}{2}},\mathellipsis,t^{N-\frac{1}{2}};q,t)\,P_{\mu^t}(t^{\frac{1}{2}}\,q^{-\lambda^{t}_{1}},\mathellipsis,t^{N-\frac{1}{2}}q^{-\lambda^{t}_{N}};q,t)\,.
\eea
Up to framing factors the RHS is exactly the S-matrix of the $SU(N)$ refined Chern-Simons theory \cite{Aganagic:2011sg}. This clearly shows that the refined Chern-Simons theory and the refined vertex are both calculating the same quantities although in different basis of symmetric functions. 

\par{In writing Eq. (\ref{pf}) we made a choice that the preferred direction of the refined topological vertex is the internal leg. For the same brane configuration instead of choosing the internal leg to be the preferred direction we could have chosen it to be along the external legs. In this case one of the branes in placed on a preferred direction and the other brane on a un-preferred one as shown in \figref{branes2}. We introduce one unit of framing for the brane corresponding to the representation $\mu$.
\begin{figure}[h]
\begin{center}
\includegraphics[scale=0.5]{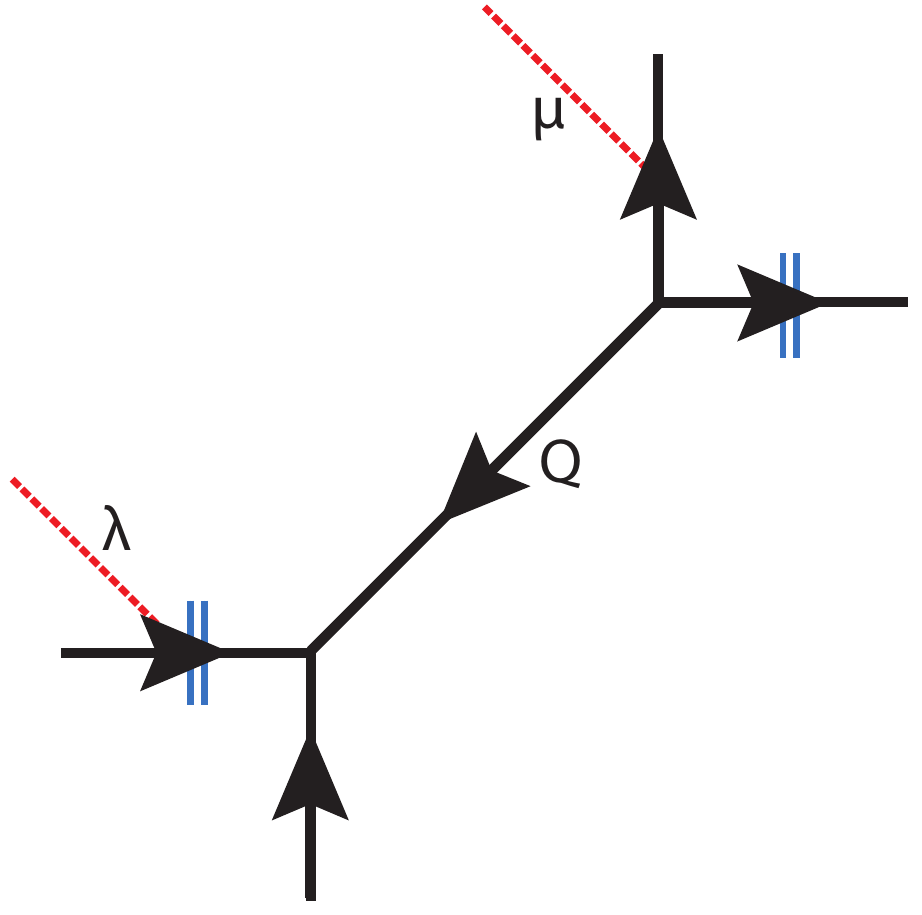}
\end{center}
\caption{The new brane configuration}\label{branes2}
\end{figure}

The corresponding refined partition function can be easily computed }
\begin{align}
Z^{(0,1)}_{\lambda\mu}&=q^{\frac{\Arrowvert\mu^{t}\Arrowvert^{2}}{2}}t^{-\frac{\Arrowvert\mu\Arrowvert^{2}}{2}}\sum_{\nu}(-Q)^{|\nu|}C_{\nu^{t}\emptyset\lambda^t}(t,q)C_{\nu\mu\emptyset}(q,t)\\ \nonumber
&=\prod_{i,j=1}^{\infty}(1-Q\, q^{-\lambda^{t}_{i}-\rho_{j}}t^{-\rho_{i}})\,q^{\frac{\Arrowvert\lambda^{t}\Arrowvert^{2}}{2}}t^{-\frac{\Arrowvert\lambda\Arrowvert^{2}}{2}}P_{\lambda^t}(t^{-\rho};q,t)s_{\mu^{t}}\Big(-t^{\rho}\sqrt{\frac{q}{t}},-Q\,q^{-\lambda^t}t^{-\rho}\Big).
\end{align}
The open amplitude can be normalized again by the closed string amplitude using the following identity
\begin{align}
\prod_{i,j=1}^{\infty}\frac{1-Q\, q^{-\lambda^{t}_{i}-\rho_{j}}t^{-\rho_{i}}}{1-Q\, q^{-\rho_{j}}t^{-\rho_{i}}}&=\prod_{(i,j)\in\lambda^t}(1-Q\,q^{-\lambda^{t}_{i}-\rho_{j}}t^{-\rho_{i}})\\ \nonumber
&=(-Q)^{|\lambda|} q^{-\frac{\Arrowvert\lambda^{t}\Arrowvert^{2}}{2}}t^{\frac{\Arrowvert\lambda\Arrowvert^{2}}{2}}\prod_{(i,j)\in\lambda^t}\left(1-Q\sqrt{\frac{q}{t}}\,q^{a'(i,j)}t^{-\ell'(i,j)}\right).
\end{align}
The normalized amplitude has a finite product rather than an infinite one
\bea\nonumber
Z_{\lambda\mu}^{(0,1)\,\mbox{\tiny{norm}}}=(-Q)^{|\lambda|}\prod_{(i,j)\in\lambda^t}\left(1-Q\sqrt{\frac{q}{t}}\,q^{a'(i,j)}t^{-\ell'(i,j)}\right)P_{\lambda^t}(t^{-\rho};q,t)\,s_{\mu^{t}}(-t^{\rho-1/2}q^{1/2},-Q\,q^{-\lambda^t}t^{-\rho}).\\
\eea
This last expression can be further simplified by recalling
\bea
P_{\lambda}(t^{-\rho},z\,t^{\rho};q,t)=P_{\lambda}(t^{-\rho};q,t)\prod_{s\in \lambda}(1-z\,q^{a'(s)}\,t^{-\ell'(s)}).
\eea
The normalized open amplitude can be written in the following compact form
\bea
Z_{\lambda\mu}^{(0,1)\,\mbox{\tiny{norm}}}=(-Q)^{|\lambda|+|\mu|}P_{\lambda^t}\Big(t^{-\rho},Q^{-1}\sqrt{\frac{q}{t}}\,t^{\rho};q,t\Big)s_{\mu^{t}}\Big(q^{-\lambda^t}t^{-\rho},Q^{-1}\sqrt{\frac{q}{t}}\,t^{\rho}\Big).
\eea
Thus for the special values of the K\"{a}hler modulus mentioned before $Q^{-1}\sqrt{\frac{q}{t}}=t^{N}$, the normalized partition function becomes
\bea\nn
Z_{\lambda\,\mu}^{(0,1)\,\mbox{\tiny{norm}}}=\Big(-\sqrt{\frac{q}{t}}\,t^{-N}\Big)^{|\lambda|+|\mu|}\,P_{\lambda^t}\Big(t^{\frac{1}{2}},t^{\frac{3}{2}},\mathellipsis, t^{N-\frac{1}{2}};q,t\Big)\,s_{\mu^t}\Big(t^{\frac{1}{2}}q^{-\lambda^{t}_{1}},t^{\frac{3}{2}}q^{-\lambda^{t}_{2}},\mathellipsis,t^{N-\frac{1}{2}}q^{-\lambda^{t}_{N}}\Big).
\eea
In terms of the holonomies $U$ and $V$ the open string partition function is given by
\bea
Z(U,V)&=&\sum_{\lambda\,\mu}Z_{\lambda\,\mu}^{(0,1)\,\mbox{\tiny{norm}}}\,P_{\lambda}(U;t,q)\,s_{\mu}(V)\\\nn
&=&\sum_{\lambda}\,\Big(-\sqrt{\frac{q}{t}}\,t^{-N}\Big)^{|\lambda|}\,P_{\lambda^t}\Big(t^{\frac{1}{2}},t^{\frac{3}{2}},\mathellipsis, t^{N-\frac{1}{2}};q,t\Big)\,P_{\lambda}(U;t,q)\prod_{i,j=1}^{N}(1-y_{i}\,\sqrt{\frac{q}{t}}\,t^{-N}\,t^{j-\frac{1}{2}}q^{-\lambda^{t}_{j}}).
\eea
If express $Z(U,V)$ in terms of Macdonald polynomials we get
\bea
Z(U,V)&=&\sum_{\lambda\,\mu}\widetilde{Z}^{(0,1)}_{\lambda\,\mu}\,P_{\lambda}(U;t,q)\,P_{\mu}(V;t,q)\\\nn
\widetilde{Z}^{(0,1)}_{\lambda\,\mu}&=&\Big(-\sqrt{\frac{q}{t}}\,t^{-N}\Big)^{|\lambda|+|\mu|}\,P_{\lambda^t}\Big(t^{\frac{1}{2}},t^{\frac{3}{2}},\mathellipsis, t^{N-\frac{1}{2}};q,t\Big)\,P_{\mu^t}\Big(t^{\frac{1}{2}}q^{-\lambda^{t}_{1}},t^{\frac{3}{2}}q^{-\lambda^{t}_{2}},\mathellipsis,t^{N-\frac{1}{2}}q^{-\lambda^{t}_{N}};q,t\Big),
\eea
which is the same as Eq. (\ref{internal}) up to framing factors.

\section{Conclusions}

We have shown that the open string partition function, corresponding to Hopf link, expressed in the basis of Macdonald polynomials gives the S-matrix of the Chern-Simons theory. This suggests that the refined Chern-Simons theory might be the underlying theory of refined topological vertex just as the Chern-Simons theory is the underlying theory of the topological vertex. It would be interesting to derive a topological vertex, following \cite{AKMV}, from the refined Chern-Simons theory and examine its relation with the refined topological vertex.

We have also argued that topological field theory of the refined Chern-Simons is the same as the Kirillov's modular tensor category based on the quantum group $U_{q}(sl_N)$ \cite{Kirillov}.  A deformation of the G/G WZW theory which seems closely related to the Kirillov's construction was  studied by Gorsky and Nekrasov \cite{GN} in the context of integrable systems. The deformed theory has features which seem closely related to the refined Chern-Simons theory \cite{progress}.

\section*{Acknowledgements}
AI would like to thank Mina Aganagic, Sergei Gukov, Nikita Nekrasov, Shamil Shakirov and Cumrun Vafa for useful discussions. AI would also like to thank the  Simon Center for Geometry and Physics for their hospitality while this work was being carried out. We would also like to thank the Simons Summer Workshop in Mathematics and Physics for their hospitality in the summer 2011.
\appendix
\numberwithin{equation}{section}
\section{Appendix A}

In this section, we want to review our notations, conventions and collect some of the useful formulas for the derivations. We want to warn the reader that the convention we follow in this note is different from the one used in \cite{Iqbal:2007ii}. The difference comes from the pictorial representation of the Young diagrams, depicted in \figref{convention}. In  \cite{Iqbal:2007ii}, for a given representation $\lambda=\{\lambda_{1},\lambda_{2},\mathellipsis \}$,we illustrated the corresponding Young diagram with columns of height $\lambda_{i}$. The arm length $a(i,j)$ and leg length $\ell(i,j)$ of a given box $(i,j)$ in the diagram were defined as the number of boxes to the right of that box and above that box, respectively. In this note we follow the more conventional notation shown on the right of \figref{convention} where $\lambda_{i}$'s are the number of boxes in the $i^{th}$ row. The arm and leg lengths are defined accordingly.

\begin{figure}[h]
\begin{center}
\includegraphics[scale=0.7]{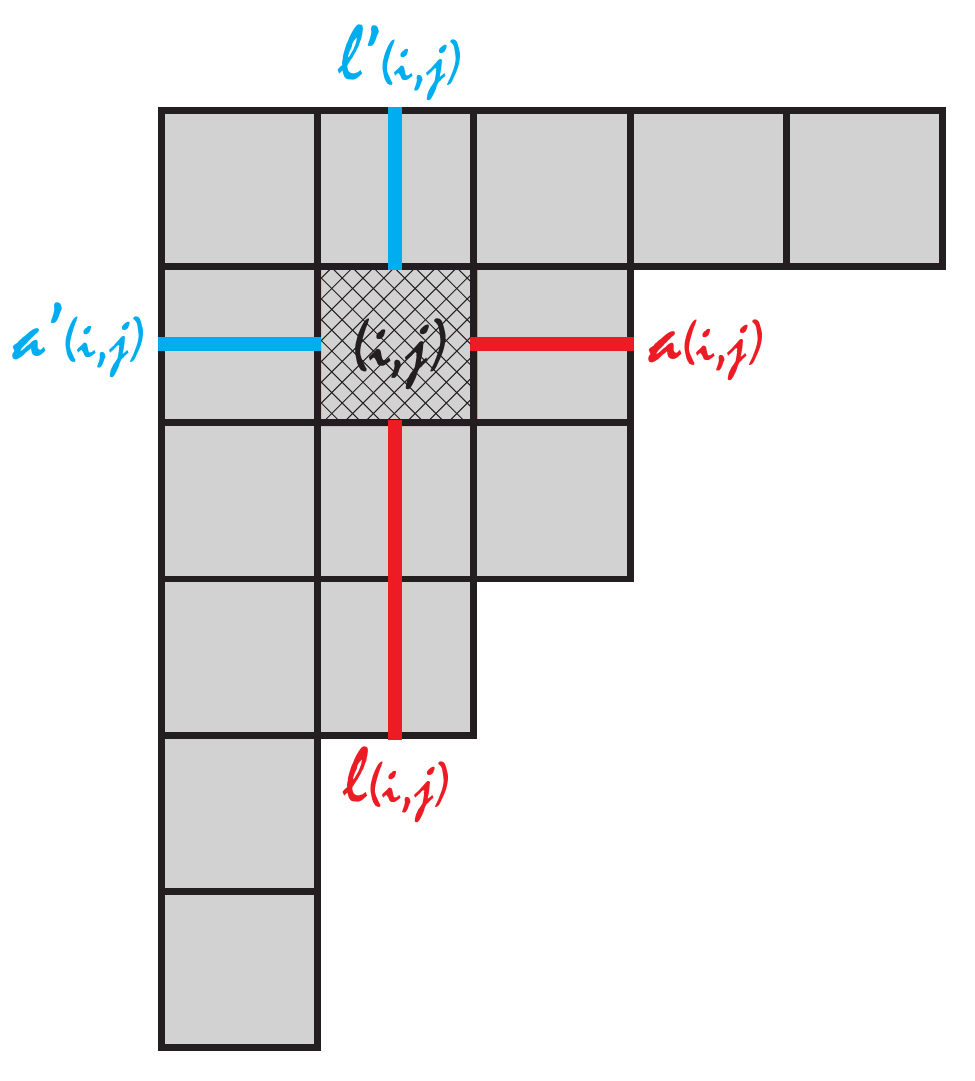}
\end{center}
\caption{The figure on the left depicts the convention used in \cite{Iqbal:2007ii}, and the figure on the right shows the present convention. $\lambda=\{5,3,3,2,1,1\}$}\label{convention}
\end{figure}
In addition to the arm and leg lengths, we define the co-arm length $a'(i,j)$ and co-leg length $\ell'(i,j)$
\begin{align}
&a(i,j)=\nu_{i}-j\,,\,\,\,\,\ell(i,j)=\nu^{t}_{j}-i\\
&a'(i,j)=j-1\,,\,\,\,\ell'(i,j)=i-1\,.
\end{align}
Note that
\bea
a_{\lambda}(i,j)=\ell_{\lambda^{t}}(j,i),\qquad\qquad\ell_{\lambda}(i,j)=a_{\lambda^{t}}(j,i)\,,
\eea
with $(i,j)\in\lambda$ and $(j,i)\in\lambda^{t}$.
The difference in the conventions reflects also on some identities used in \cite{Iqbal:2007ii}, here we use
\begin{align}
&n(\lambda)=\sum_{i=1}^{\ell(\lambda)}(i-1)\lambda_{i}=\frac{1}{2}\sum_{i=1}^{\ell(\lambda)}\lambda^{t}_{i}(\lambda^{t}_{i}-1)=\sum_{(i,j)\in\lambda}\ell(i,j)=\sum_{(i,j)\in\lambda}\ell'(i,j)=\frac{\Arrowvert\lambda^{t}\Arrowvert^{2}}{2}-\frac{\lambda}{2}\,,\\
&n(\lambda^{t})=\sum_{i=1}^{\ell(\lambda^{t})}(i-1)\lambda^{t}_{i}=\frac{1}{2}\sum_{i=1}^{\ell(\lambda^{t})}\lambda_{i}(\lambda_{i}-1)=\sum_{(i,j)\in\lambda}a(i,j)=\sum_{(i,j)\in\lambda}a'(i,j)=\frac{\Arrowvert\lambda\Arrowvert^{2}}{2}-\frac{\lambda}{2}\,,
\end{align}
with $\ell(\lambda)$ being the number of non-zero $\lambda_{i}$'s, or in other words, it is the number of rows in the Young diagram. The hook length $h(i,j)$ and the content $c(i,j)$ are defined as
\bea
h(i,j)=a(i,j)+\ell(i,j)+1,\qquad c(i,j)=j-i\,,
\eea
which satisfy
\begin{align}
&\sum_{(i,j)\in\lambda}h(i,j)=n(\lambda^{t})+n(\lambda)+|\lambda|,\\
&\sum_{(i,j)\in\lambda}c(i,j)=n(\lambda^{t})-n(\lambda).
\end{align}
The refined topological vertex in this convention is given by
\bea\nn
C_{\lambda\,\mu\,\nu}(t,q)&=&\Big(\frac{q}{t}\Big)^{\frac{||\mu||^2}{2}}\,t^{\frac{\kappa(\mu)}{2}}\,q^{\frac{||\nu||^2}{2}}\,\widetilde{Z}_{\nu}(t,q)
\sum_{\eta}\Big(\frac{q}{t}\Big)^{\frac{|\eta|+|\lambda|-|\mu|}{2}}\,s_{\lambda^{t}/\eta}(t^{-\rho}\,q^{-\nu})\,s_{\mu/\eta}(t^{-\nu^t}\,q^{-\rho})\,,\\
\eea
where $s_{\lambda/\eta}(\mathbf{x})$ is the skew-Schur function, $\rho=\{-\frac{1}{2},-\frac{3}{2},-\frac{5}{2},\cdots\}$ and $\widetilde{Z}_{\nu}(t,q)$ is given by
\bea
\widetilde{Z}_{\nu}(t,q)&=&\prod_{(i,j)\in \nu}\Big(1-q^{a(i,j)}\,t^{\ell(i,j)+1}\Big)^{-1},
\eea
which is related to the Macdonald polynomials
\bea
t^{\frac{\Arrowvert\nu^t\Arrowvert^2}{2}}\,\widetilde{Z}_{\nu}(t,q)&=&P_{\nu}(t^{-\rho};q,t),\\
q^{\frac{\Arrowvert\nu\Arrowvert^2}{2}}\,\widetilde{Z}_{\nu^t}(q,t)&=&P_{\nu^t}(q^{-\rho};t,q).
\eea
For the Macdonald polynomials with this special set of argument we have
\bea
P_{\lambda}(t^{\rho};q,t)=(-1)^{|\lambda|}q^{n(\lambda^{t})}t^{-n(\lambda)}P_{\lambda}(t^{-\rho};q,t).
\eea
The Macdonald polynomials also satisfy
\bea
P_{\lambda}(\mathbf{x};q,t)=P_{\lambda}(\mathbf{x};q^{-1},t^{-1})
\eea
For completeness, let us also review some identities regarding to Schur functions and MacDonald polynomials:
\begin{align}
&s_{\lambda}(\mathbf{x})s_{\mu}(\mathbf{x})=\sum_{\eta}N_{\lambda\mu}^{\eta}s_{\eta}(\mathbf{x})\,,\label{schurfactor}\\
&s_{\lambda/\mu}(\mathbf{x})=\sum_{\eta}N^{\lambda}_{\mu\eta}s_{\eta}(\mathbf{x})\,,\\
&P_{\lambda}(\mathbf{x};q,t)P_{\mu}(\mathbf{x};q,t)=\sum_{\eta}\widehat{N}^{\eta}_{\lambda\mu}P_{\eta}(\mathbf{x};q,t)\,,\label{macdonald}\\
&s_{\eta}(\mathbf{x})=\sum_{\sigma}U_{\eta\sigma}P_{\sigma}(\mathbf{x};q,t)\,.\label{schurmacdonald}
\end{align}
The following relations among Macdonald polynomials prove to be very useful for our computations (Page 332, Eq(6.6) of \cite{macdonald})
\bea\label{ratio}
\frac{P_{\mu}(q^{-\lambda}\,t^{-\rho};q,t)}{P_{\mu}(t^{-\rho};q,t)}&=&\frac{P_{\lambda}(q^{-\mu}\,t^{-\rho};q,t)}{P_{\lambda}(t^{-\rho};q,t)}\,,\\
P_{\sigma}(t^{-\rho},z\,t^{\rho};q,t)&=&P_{\sigma}(t^{-\rho};q,t)\prod_{s\in \sigma}(1-z\,q^{a'(s)}\,t^{-\ell'(s)})\,.\label{product}
\eea
The following sum rules are essential for vertex computations
\begin{align}
&\sum_{\eta}s_{\eta/\lambda}(\mathbf{x})s_{\eta/\mu}(\mathbf{y})=\prod_{i,j=1}^{\infty}(1-x_{i}y_{j})^{-1}\sum_{\tau}s_{\mu/\tau}(\mathbf{x})s_{\lambda/\tau}(\mathbf{y})\,.\\
&\sum_{\eta}s_{\eta^{t}/\lambda}(\mathbf{x})s_{\eta/\mu}(\mathbf{y})=\prod_{i,j=1}^{\infty}(1+x_{i}y_{j})\sum_{\tau}s_{\mu^{t}/\tau}(\mathbf{x})s_{\lambda^{t}/\tau^{t}}(\mathbf{y})\,,\\
&\sum_{\eta}P_{\eta}(\mathbf{x};q,t)P_{\eta^{t}}(\mathbf{x};t,q)=\prod_{i,j=1}^{\infty}(1+x_{i}y_{j})\,.
\end{align}
We have considered the normalized amplitudes, both the open and closed amplitudes are infinite series in the K\"{a}hler parameters, however, their ratio is finite as a result of the following identity
\bea\label{infinite}
\prod_{i,j=1}^{\infty}\frac{1-Q\,q^{-\lambda_{i}+j-1/2}t^{i-1/2}}{1-Q\,q^{j-1/2}t^{i-1/2}}=\prod_{s\in\lambda}\left (1-Q\sqrt{\frac{t}{q}}\, q^{-a'(s)}t^{\ell(s)}\right).
\eea
\section{Appendix B}

In this section, we want to prove the identity Eq. (\ref{identityx}) we used to show the equivalence of the open topological string amplitudes for two different choices of the preferred direction of the refined topological vertex. Our derivation relies on the results of \cite{CM}. They showed that the normalized one- and two-point function of the Macdonald polynomials with measure $\Delta_{q,t}$ times a Gaussian are given by \cite{CM}
\begin{align}
&\langle P_{\lambda}(\mathbf{x};q,t) \rangle=t^{-\frac{|\lambda|}{2}}q^{\frac{\Arrowvert\lambda\Arrowvert^{2}}{2}}t^{-\frac{{\Arrowvert\lambda^{t}\Arrowvert^{2}}}{2}}P_{\lambda}(1,t,\mathellipsis,t^{N-1};q,t)\,,\\
&\langle P_{\lambda}(\mathbf{x};q,t)\, P_{\mu}(\mathbf{x};q,t) \rangle=t^{N|\mu|}t^{\frac{|\mu|}{2}-\frac{|\lambda|}{2}}q^{\frac{\Arrowvert\lambda\Arrowvert^{2}+\Arrowvert\mu\Arrowvert^{2}}{2}}t^{-\frac{\Arrowvert\lambda^{t}\Arrowvert^{2}+\Arrowvert\mu^{t}\Arrowvert^{2}}{2}}\\\nn
&\qquad\qquad\qquad\qquad\qquad \times P_{\lambda}(1,t,\mathellipsis,t^{N-1};q,t)\, P_{\mu}(q^{\lambda_{1}},q^{\lambda_{2}}t^{-1},\mathellipsis,q^{\lambda_{N}}t^{-(N-1)};q,t)\,.
\end{align}
By Eq. \ref{macdonald} we know that these correlators  are related to each other.
\bea\nn
t^{N|\mu|}t^{\frac{|\mu|}{2}-\frac{|\lambda|}{2}}q^{\frac{\Arrowvert\lambda\Arrowvert^{2}+\Arrowvert\mu\Arrowvert^{2}}{2}}t^{-\frac{\Arrowvert\lambda^{t}\Arrowvert^{2}+\Arrowvert\mu^{t}\Arrowvert^{2}}{2}}\,&P_{\lambda}(t,t^{2},\mathellipsis,t^{N};q,t)\, P_{\mu}(q^{\lambda_{1}}t^{-1},q^{\lambda_{2}}t^{-2},\mathellipsis,q^{\lambda_{N}}t^{-N};q,t)\\
&=\sum_{\eta}\widehat{N}^{\eta}_{\lambda\mu}t^{-\frac{|\eta|}{2}}q^{\frac{\Arrowvert\eta\Arrowvert^{2}}{2}}t^{-\frac{{\Arrowvert\eta^{t}\Arrowvert^{2}}}{2}}P_{\eta}(t,t^{2},\mathellipsis,t^{N};q,t)\,.
\eea
This last expression is not in the form we used above, however, it is straightforward to manipulate this identity into the desired form. Let us multiply both sides by $t^{-N(|\lambda|+|\mu|)}$, and note that $\widehat{N}^{\eta}_{\lambda\mu}$ vanishes unless $|\eta|=|\lambda|+|\mu|$ is satisfied. Then we get
\bea\nn
q^{\frac{\Arrowvert\lambda\Arrowvert^{2}+\Arrowvert\mu\Arrowvert^{2}}{2}}t^{-\frac{\Arrowvert\lambda^{t}\Arrowvert^{2}+\Arrowvert\mu^{t}\Arrowvert^{2}}{2}}\,&P_{\lambda}(t^{-1/2},t^{-3/2},\mathellipsis,t^{-N+1/2};q,t)\, P_{\mu}(q^{\lambda_{1}}t^{-1/2},q^{\lambda_{2}}t^{-3/2},\mathellipsis,q^{\lambda_{N}}t^{-N+1/2};q,t)\\
&=\sum_{\eta}\widehat{N}^{\eta}_{\lambda\mu}q^{\frac{\Arrowvert\eta\Arrowvert^{2}}{2}}t^{-\frac{{\Arrowvert\eta^{t}\Arrowvert^{2}}}{2}}P_{\eta}(t^{-1/2},t^{-3/2},\mathellipsis,t^{-N+1/2};q,t)\,.
\eea
Taking $(q,t)\mapsto (q^{-1},t^{-1})$ and keeping in mind that $P_{\lambda}({\bf x};q^{-1},t^{-1})=P_{\lambda}({\bf x};q,t)$ and $\widehat{N}^{\nu}_{\lambda\,\mu}(q^{-1},t^{-1})=\widehat{N}^{\nu}_{\lambda\,\mu}(q,t)$ we get Eq. (\ref{identityx}).

\section{Appendix C}
\par{In this section, we want to present the details of the derivation of Eq. (\ref{norm1}). We will ignore the factors related to framing for the notational ease. Let us begin with Eq. (\ref{xcv}):}
\begin{align}
\widetilde{Z}^{(1,0)}_{\lambda\mu}=\sum_{\nu}(-Q)^{|\nu|}P_{\nu}(t^{-\rho};q,t)P_{\nu^{t}}(q^{-\rho};t,q)P_{\lambda}(q^{-\nu}t^{-\rho};q,t)P_{\mu}(q^{-\nu}t^{-\rho};q,t)\,.
\end{align}
Note that the product of two Macdonald polynomials with the same arguments can be expanded in Macdonald polynomials (Eq. (\ref{macdonald})) therefore
\bea
\widetilde{Z}^{(1,0)}_{\lambda\mu}=\sum_{\nu,\sigma}(-Q)^{|\nu|}\widehat{N}^{\sigma}_{\lambda\,\mu}\,P_{\nu^{t}}(q^{-\rho};t,q)P_{\nu}(t^{-\rho};q,t)P_{\sigma}(t^{-\rho}q^{-\nu};q,t)\,.
\eea
The $\nu$-sum can be explicitly performed after using Eq. \ref{ratio}, the amplitude becomes
\bea
\widetilde{Z}^{(1,0)}_{\lambda\mu}=\sum_{\sigma}\widehat{N}^{\sigma}_{\lambda\,\mu} P_{\sigma}(t^{-\rho};q,t)\prod_{i,j=1}^{\infty} \left(1-Q\, q^{-\sigma_{i}-\rho_{j}}t^{-\rho_{i}}\right).
\eea
Let us finally normalize by the closed amplitude (Eq. \ref{infinite})
\bea
\widetilde{Z}^{(1,0)\,\mbox{\tiny{norm}}}_{\lambda\mu}=\sum_{\sigma}\widehat{N}^{\sigma}_{\lambda\,\mu}\, P_{\sigma}(t^{-\rho};q,t)\prod_{(i,j)\in\sigma}\left (1-Q\,q^{-j+\frac{1}{2}}t^{i-\frac{1}{2}} \right).
\eea
This last expression can be even further simplified using Eq. (\ref{product}) to get the result we have given in the main text:
\bea
\widetilde{Z}_{\lambda\mu}^{(1,0)\,\mbox{\tiny{norm}}}=(-Q)^{|\lambda|+|\mu|}\sum_{\sigma}\widehat{N}^{\sigma}_{\lambda\,\mu}\, t^{\frac{\Arrowvert\sigma^{t}\Arrowvert^{2}}{2}}q^{-\frac{\Arrowvert\sigma\Arrowvert^{2}}{2}}P_{\sigma}(t^{-\rho},Q^{-1}\sqrt{\frac{q}{t}}\,t^{\rho};q,t)\,.
\eea

\bibliography{references}

\providecommand{\href}[2]{#2}\begingroup\raggedright\begin{thebibliography}{10}

\bibitem{Witten:1988hf}
E.~Witten, ``{Quantum field theory and the Jones polynomial},'' {\em Commun.
  Math. Phys.} {\bf 121} (1989)
351.

\bibitem{Khovanov}
M.~Khovanov, ``{A categorification of the Jones ploynomial},''
\href{http://arXiv.org/abs/math.QA/9908171}{{\tt math.QA/9908171}}.

\bibitem{Ooguri:1999bv}
H.~Ooguri and C.~Vafa, ``{Knot invariants and topological strings},'' {\em
  Nucl. Phys.} {\bf B577} (2000) 419--438,
\href{http://arXiv.org/abs/hep-th/9912123}{{\tt hep-th/9912123}}.

\bibitem{Gukov:2004hz}
S.~Gukov, A.~S. Schwarz, and C.~Vafa, ``{Khovanov-Rozansky homology and
  topological strings},'' {\em Lett. Math. Phys.} {\bf 74} (2005) 53--74,
\href{http://arXiv.org/abs/hep-th/0412243}{{\tt hep-th/0412243}}.

\bibitem{Nekrasov:2002qd}
N.~A. Nekrasov, ``{Seiberg-Witten prepotential from instanton counting},'' {\em
  Adv. Theor. Math. Phys.} {\bf 7} (2004) 831--864,
\href{http://arXiv.org/abs/hep-th/0206161}{{\tt hep-th/0206161}}.

\bibitem{Hollowood:2003cv}
T.~J. Hollowood, A.~Iqbal, and C.~Vafa, ``{Matrix Models, Geometric Engineering
  and Elliptic Genera},'' {\em JHEP} {\bf 03} (2008) 069,
\href{http://arXiv.org/abs/hep-th/0310272}{{\tt hep-th/0310272}}.

\bibitem{Iqbal:2007ii}
A.~Iqbal, C.~Kozcaz, and C.~Vafa, ``{The Refined topological vertex},'' {\em
  JHEP} {\bf 0910} (2009) 069, \href{http://arXiv.org/abs/hep-th/0701156}{{\tt
  hep-th/0701156}}.

\bibitem{Gukov:2007tf}
S.~Gukov, A.~Iqbal, C.~Kozcaz, and C.~Vafa, ``{Link homologies and the refined
  topological vertex},'' {\em Commun. Math. Phys.} {\bf 298} (2010) 757--785,
\href{http://arXiv.org/abs/0705.1368}{{\tt 0705.1368}}.

\bibitem{Aganagic:2011sg}
M.~Aganagic and S.~Shakirov, ``{Knot Homology from Refined Chern-Simons
  Theory},''
\href{http://arXiv.org/abs/arXiv:1105.5117}{{\tt arXiv:1105.5117}}.

\bibitem{Kozcaz:2010af}
C.~Kozcaz, S.~Pasquetti, and N.~Wyllard, ``{A \& B model approaches to surface
  operators and Toda theories},'' {\em JHEP} {\bf 08} (2010) 042,
\href{http://arXiv.org/abs/1004.2025}{{\tt 1004.2025}}.

\bibitem{Dimofte:2010tz}
T.~Dimofte, S.~Gukov, and L.~Hollands, ``{Vortex Counting and Lagrangian
  3-manifolds},''
\href{http://arXiv.org/abs/arXiv:1006.0977}{{\tt arXiv:1006.0977}}.

\bibitem{Iqbal:2008ra}
A.~Iqbal, C.~Kozcaz, and K.~Shabbir, ``{Refined Topological Vertex, Cylindric
  Partitions and the U(1) Adjoint Theory},'' {\em Nucl.Phys.} {\bf B838} (2010)
  422--457, \href{http://arXiv.org/abs/0803.2260}{{\tt 0803.2260}}.

\bibitem{Awata:2009sz}
H.~Awata and H.~Kanno, ``{Macdonald operators and homological invariants of the
  colored Hopf link},'' {\em J. Phys.} {\bf A44} (2011) 375201,
\href{http://arXiv.org/abs/0910.0083}{{\tt 0910.0083}}.

\bibitem{macdonald}
I.~G. Macdonal, ``Symmetric functions and hall polynomials,'' {\em Oxford
  Mathematical Monographs, Oxford Science Publications}
(second edition, 1995).

\bibitem{Kirillov}
P.~I. {Etingof} and A.~A. {Kirillov}, Jr., ``{Representation-theoretic proof of
  the inner product and symmetry identities for MacDonald's polynomials},''
  \href{http://arXiv.org/abs/arXiv:math/9411232}{{\tt arXiv:math/9411232}}.

\bibitem{Moore:1988uz}
G.~W. Moore and N.~Seiberg, ``{Polynomial Equations for Rational Conformal
  Field Theories},'' {\em Phys. Lett.} {\bf B212} (1988)
451.

\bibitem{Moore:1988qv}
G.~W. Moore and N.~Seiberg, ``{Classical and Quantum Conformal Field Theory},''
  {\em Commun.Math.Phys.} {\bf 123} (1989) 177.

\bibitem{Elitzur:1989nr}
S.~Elitzur, G.~W. Moore, A.~Schwimmer, and N.~Seiberg, ``{Remarks on the
  Canonical Quantization of the Chern-Simons- Witten Theory},'' {\em Nucl.
  Phys.} {\bf B326} (1989)
108.

\bibitem{Kirillov2}
A.~{Kirillov}, Jr., ``{On inner product in modular tensor categories. I},''
  \href{http://arXiv.org/abs/q-alg/9508017}{{\tt q-alg/9508017}}.

\bibitem{Kirillov3}
B.~Bakalov and A.~A. Kirillov~Jr., {\em Lectures on Tensor Categories and
  Modular Functor}.
\newblock American Mathematical Society, 2001.

\bibitem{Rowell}
E.~C. {Rowell}, ``{From Quantum Groups to Unitary Modular Tensor Categories},''
  \href{http://arXiv.org/abs/arXiv:math/0503226}{{\tt arXiv:math/0503226}}.

\bibitem{AKMV}
M.~Aganagic, A.~Klemm, M.~Marino, and C.~Vafa, ``{The Topological Vertex},''
  {\em Commun. Math. Phys.} {\bf 254} (2005) 425--478,
\href{http://arXiv.org/abs/hep-th/0305132}{{\tt hep-th/0305132}}.

\bibitem{GN}
A.~Gorsky and N.~Nekrasov, ``{Relativistic Calogero-Moser model as gauged WZW
  theory},'' {\em Nucl. Phys.} {\bf B436} (1995) 582--608,
\href{http://arXiv.org/abs/hep-th/9401017}{{\tt hep-th/9401017}}.

\bibitem{progress}
``work in progress,''.

\bibitem{CM}
P.~{Etingof} and A.~{Kirillov}, Jr, ``{On Cherednik-Macdonald-Mehta
  identities},'' \href{http://arXiv.org/abs/q-alg/9712051}{{\tt
  q-alg/9712051}}.

\end{thebibliography}\endgroup

\end{document}